\documentclass{article}
\pdfoutput=1

\usepackage{authblk}                    
\usepackage{graphicx,epsfig,epstopdf}   
\usepackage{booktabs}                   
\usepackage{amsmath,amsthm,amssymb}     
\usepackage{color}                      
\usepackage{cite}                       
\usepackage[table]{xcolor}              
\usepackage{siunitx}                 
\usepackage{nicematrix}
\usepackage{multirow} 

\usepackage{float}

\DeclareUnicodeCharacter{2212}{-}

\graphicspath{ {figures/} }

\usepackage[paperwidth = 17cm,          
paperheight = 24cm,
left = 1.75cm,
right = 1.75cm,
top = 2cm,
bottom = 2cm]{geometry}

\usepackage[
    bookmarks         = true,
    bookmarksnumbered = true,
    bookmarksopen     = true,
    colorlinks        = true,
    linktocpage       = true,%
    urlcolor          = blue,
    linkcolor         = blue,
    citecolor         = red,
    ]{hyperref}%

\definecolor{coolblack}{rgb}{0.0, 0.18, 0.39}

\title{\textbf{Gravitational waves from high-temperature vacuum decay in scale-invariant models: nanohertz vs. millihertz regimes}}
\author[1]{Ahmad Mohamadnejad\thanks{mohamadnejad.a@lu.ac.ir}}
\affil[1]{Department of Physics, Lorestan University,
Khorramabad, Iran}

\begin{document}

\maketitle

\begin{abstract}
We present a comprehensive analysis of high-temperature vacuum decay and the resulting stochastic gravitational wave (GW) background within the framework of general scale-invariant models. The effective potential is constructed to include tree-level contributions, the Coleman-Weinberg correction, finite-temperature effects, and the Daisy resummation technique, culminating in a high-temperature form. We investigate the dynamics of the first-order phase transition, calculating the critical and nucleation temperatures, the supercooling parameter, and the key transition parameters $\alpha$ (transition strength) and $\beta$ (inverse duration). The vacuum decay is found to be dominated by sphaleron transitions rather than quantum tunneling. We compute the full GW spectrum arising from bubble collisions, sound waves, and turbulence. An extensive numerical scan reveals two distinct phenomenological regimes: one produces nanohertz-frequency GW signals potentially detectable by Pulsar Timing Arrays (PTA), while the other yields millihertz-frequency signals that are prime targets for future space-based interferometers like the Laser Interferometer Space Antenna (LISA).
\end{abstract}

\noindent\hrulefill
\tableofcontents
\noindent\hrulefill

\numberwithin{equation}{section}

\section{Introduction} \label{sec1}
"The concept of spontaneous symmetry breaking is a cornerstone of modern theoretical physics, providing the mechanism for generating particle masses in the Standard Model (SM) through the Higgs mechanism \cite{Higgs:1964pj, Englert:1964et, Guralnik:1964eu}. While phenomenologically successful, the Higgs sector of the SM introduces significant theoretical puzzles, notably the hierarchy problem \cite{'tHooft:1979bh}. Scale-invariant theories offer a compelling framework to address this issue by positing that the fundamental laws of physics are invariant under scale transformations, with all mass scales, including the electroweak scale, generated dynamically through dimensional transmutation \cite{Coleman:1973jx, Gildener:1976ih}. This approach has seen a significant revival in recent years as a paradigm for physics beyond the SM \cite{Foot:2007ay, Hur:2011sv, Farzinnia:2013pga, Alexander-Nunneley:2010tyr, Karam:2015jta, Karam:2016rsz, Guo:2015lxa, YaserAyazi:2018lrv, YaserAyazi:2019caf, Mohamadnejad:2019wqb, Mohamadnejad:2019vzg, Mohamadnejad:2021tke, Hosseini:2023qwu, Abkenar:2024ket, Abkenar:2025urg, Ahmed:2025gww, Aoki:2024jhr, Ahriche:2023jdq}.

A critical consequence of these models is the prediction of a strong first-order phase transition (SFOPT) in the early universe as it cools and transitions from a symmetric high-temperature phase to a symmetry-broken vacuum. Such a phase transition is a rich source of phenomenology. It can provide a viable pathway for electroweak baryogenesis, potentially explaining the observed matter-antimatter asymmetry \cite{Kuzmin:1985mm, Morrissey:2012db}. Furthermore, a SFOPT generates a stochastic background of GWs through the violent dynamics of bubble nucleation, expansion, and collision \cite{Kosowsky:1992vn, Caprini:2015zlo}. For a comprehensive and pedagogical review of gravitational waves from cosmological phase transitions, we refer the reader to \cite{Athron:2023xlk}. Recent studies have also highlighted the model-independent features of phase transitions and GW production in scale-invariant settings \cite{Salvio:2023qgb, Salvio:2023ynn, Salvio:2023blb}, while analyses of theoretical uncertainties in such calculations can be found in \cite{Athron:2023rfq}.

The detection of this primordial GW signal would provide a unique, uncontaminated window into the physics of the early universe and fundamental particle interactions at high energies, far beyond the direct reach of terrestrial colliders \cite{Caprini:2019egz}. Upcoming GW observatories, such as the Laser Interferometer Space Antenna (LISA) \cite{LISA2017, Baker:2019nia}, offer the exciting prospect of probing these cosmic phase transitions, complementing efforts with Pulsar Timing Arrays (PTA) \cite{NANOGrav:2023gor} and other ground-based and space-based detectors.

In this work, we undertake a model-independent study of vacuum decay and GW production in a general class of scale-invariant models at high temperatures. We derive the general form of the high-temperature effective potential and analyze the dynamics of the phase transition, including the calculation of the sphaleron-dominated decay rate. Our primary focus is to compute the resultant GW spectrum and map the model parameters to observable signatures. We identify distinct regimes in the parameter space, delineating scenarios that are either ruled out by cosmological constraints or present promising targets for future detection.

The paper is organized as follows: In Section~\ref{sec2}, we construct the effective potential. Section~\ref{sec3} is devoted to the analysis of metastable vacuum decay and phase transition dynamics. The numerical results for the GW spectrum from various sources are presented and discussed in Section~\ref{sec4}. Finally, we conclude in Section~\ref{sec5}."

\section{The Effective potential} \label{sec2}
In quantum field theory, the effective potential captures the impact of quantum fluctuations on classical potential. It plays a key role in understanding concepts such as vacuum stability, symmetry breaking, and cosmological phase transitions. In this study, we will concentrate on the effective potential within scale-invariant models, emphasizing tree-level contributions, the Coleman-Weinberg correction, finite-temperature effects, and the Daisy resummation technique.

\subsection{Tree level potential}
At the tree level, the classical scale-invariant action can be extended to accommodate any number of scalar fields. The scale-invariant actions must maintain invariance under the rescaling of both the fields and the spacetime coordinates. For a model to be truly scale invariant, the action should be invariant under the transformation
\begin{equation}
x^{\mu} \to \omega^{-1} x^{\mu}, \quad \phi_i \to \omega \phi_i, \quad \text{for any parameter } \omega > 0.
\label{2-1}
\end{equation}
This means that the potential must be constructed from homogeneous function terms such that the total degree of each term in the potential is the same and composed solely of quartic ones.
For a collection of \( n \) scalar fields, represented as \( \phi_1, \phi_2, \ldots, \phi_n \), the most general form of the potential in scale-invariant models is given by
\begin{equation}  
V^{tree}( \phi_1, \phi_2, \ldots, \phi_n ) = \frac{1}{4!} \sum_{i,j,k,l}^{n} \lambda_{ijkl} \phi_i \phi_j \phi_k \phi_l, \label{2-2}
\end{equation}
where $ \lambda_{ijkl} $ are the dimensionless coupling constants of the model. Within the field space, there exists a direction known as the flat direction, represented by the unit vector \( \mathbf{N}_0 = (N_1, N_2, \ldots, N_n) \), along which the potential exhibits a non-trivial minimum:
\begin{equation}  
V^{tree}( N_1 \phi, N_2 \phi, \ldots, N_n \phi ) = \frac{\phi^4}{4!} \sum_{i,j,k,l}^{n} \lambda_{ijkl} N_i N_j N_k N_l = 0.
\label{2-3}
\end{equation}
For the potential (\ref{2-2}) to have a local minimum along the flat direction \( \mathbf{N}_0 \), the following two conditions must be satisfied
\begin{enumerate}
\item First-order condition (vanishing gradient): The gradient of the potential must be zero along the flat direction \( \mathbf{N}_0 \):
\begin{equation}  
 \nabla V^{tree} = \left( \frac{\partial V^{tree}}{\partial \phi_1}, \frac{\partial V^{tree}}{\partial \phi_2}, ..., \frac{\partial V^{tree}}{\partial \phi_n} \right) \bigg|_{\mathbf{N}_0} = \mathbf{0}.
\label{2-4}
\end{equation}
This implies that all first partial derivatives of the potential with respect to each scalar field must be equal to zero along the flat direction \( \mathbf{N}_0 \).
\item Second-order condition (positive definite Hessian): The Hessian matrix \( H \) of the potential, which is the matrix of second derivatives, must be positive definite or zero along the flat direction \( \mathbf{N}_0 \). The Hessian is defined as
\begin{equation}  
H_{ij} = \frac{\partial^2 V^{tree}}{\partial \phi_i \partial \phi_j} \bigg|_{\mathbf{N}_0} .
\label{2-5}
\end{equation}
The condition for the Hessian to be positive definite or zero involves checking that all leading principal minors of \( H \) are not negative, or equivalently, that the eigenvalues of \( H \) are all either positive or zero.
\end{enumerate}
These conditions make some constraints on the couplings $ \lambda_{ijkl} $ ensuring that \( \mathbf{N}_0 \) is indeed a local minimum of the potential. Keep in mind that the field along the flat direction remains massless, whereas all other fields can have mass at the tree level.

\subsection{One loop correction}
The Coleman-Weinberg potential~\cite{Coleman:1973jx} accounts for radiative corrections that arise from quantum fluctuations in the field. The effective potential including 1-loop corrections can be expressed as
\begin{equation} 
V_{1}^{\rm CW} =  \frac{1}{64 \pi^{2}}  \sum_{k} g_{k}  M_{k}^{4}  \left(   \ln \frac{M_{k}^{2}}{\Lambda^{2}} - C_{k} \right) . \label{2-6}
\end{equation}
In this equation, \( C_{k}=3/2 \) for scalar and spinor fields and \( C_{k}=5/6 \) for vector fields, and $ \Lambda $ represents the renormalization scale. The quantity \( M_{k} \) represents the tree-level, field-dependent mass of particle \( k \), while \( g_{k} \) indicates the number of degrees of freedom, defined as
\begin{equation}
g_{k} =  (-1)^{2 s_{k}} q_{k} N_{k} (2 s_{k} + 1), \label{2-7}
\end{equation}
where \( s_{k} \) represents the spin, \( N_{k} \) denotes the number of colors, and \( q_{k}=1 \) for neutral particles and \( q_{k}=2 \) for charged particles.

To ensure a real value for Coleman-Weinberg potential, the field-dependent mass squared for the particles cannot be negative. Consequently, the field space must be restricted to the flat direction. Along this direction, \( M_{\phi} = 0 \), meaning that we do not account for the contribution of the scalon field in the Coleman-Weinberg potential. Additionally, Goldstone bosons are massless along the flat direction, so they don’t influence the minimum of the tree-level potential, and their field-dependent masses are also omitted.

Importantly, along the flat direction, we can replace all field-dependent mass contributions with \( M_{k} \rightarrow \frac{m_{k}}{\nu} \phi \), where \( m_{k} \) denotes the measured mass of particle \( k \). Applying this substitution to Eq. (\ref{2-6}) yields the well-known Gildener-Weinberg formula \cite{Gildener:1976ih}:
\begin{equation}
V^{\rm GW} = a \phi^{4} + b \phi^{4} \ln \frac{\phi^{2}}{\Lambda^{2}} , \label{2-8}
\end{equation}
with \( a \) and \( b \) defined as
\begin{align}
&a =  \frac{1}{64 \pi^{2} \nu^{4}}  \sum_{k} g_{k}  m_{k}^{4} \left(  \ln \frac{m_{k}^{2}}{\nu^{2}} - C_{k}  \right)   , \nonumber\\
&b = \frac{1}{64 \pi^{2} \nu^{4}} \sum_{k} g_{k}  m_{k}^{4} . \label{2-9}
\end{align}

Since, along the flat direction, \( V_{tree} = 0 \), therefore, to locate the true vacuum, one must find the minimum of the one-loop potential given in Eq. (\ref{2-8}), which is:
\begin{equation}
\langle \phi \rangle = \nu = \Lambda e^{-(\frac{a}{2b} + \frac{1}{4})} .  \label{2-10}
\end{equation}
By combining Eq. (\ref{2-10}) with Eq. (\ref{2-8}), we can eliminate the RG scale \( \Lambda \) and derive a simplified expression for the one-loop potential in terms of the true vacuum expectation value \( \nu \) and the coefficient \( b \)
\begin{equation}
V^{\rm GW} = b \phi^{4} \left(  \ln \frac{\phi^{2}}{\nu^{2}} - \frac{1}{2} \right) .  \label{2-11}
\end{equation}
From this equation, we can determine the mass squared of the scalar field \( m_{\phi}^{2} \) as follows:
\begin{equation}
m_{\phi}^{2} = \frac{d^2 V^{\rm GW}}{d \phi^{2}} \bigg\rvert_{\nu} =8b \nu^2.  \label{2-12}
\end{equation} 

\subsection{High temperature contribution}
Now we consider the finite-temperature 1-loop effective potential, which allows us to calculate the vacuum expectation values of scalar fields in the presence of a thermal bath at temperature \( T \). The 1-loop corrections at finite temperature are expressed as \cite{Dolan:1973qd}  
\begin{equation}  
V^{T}_{1} = \frac{T^{4}}{2 \pi^{2}}  \sum_{k} g_{k} J_{\text{B,F}} \left( \frac{M_{k}}{T} \right), \label{2-13}  
\end{equation}  
where the thermal functions for bosons and fermions are defined by  
\begin{equation}  
J_{\text{B,F}}(x) =  \int_{0}^{\infty} dy \, y^{2} \ln \left(1 \mp e^{- \sqrt{y^{2}+x^{2}}} \right). \label{2-14}  
\end{equation}  
As the temperature approaches zero, these functions vanish. Although these integrals cannot be expressed in terms of standard functions, they can be numerically evaluated relatively easily and approximated in different regimes. For instance, in the high-temperature limit (\( x \ll 1 \)), both \( J_{\text{B}}(x) \) and \( J_{\text{F}}(x) \) have convenient closed forms \cite{Dolan:1973qd}
\begin{align}  
&J_{\text{B}}^{\text{high-T}}(x) = -\frac{\pi ^4}{45} + \frac{\pi ^2 }{12} x^2 - \frac{\pi}{6} x^3 - \frac{1}{32} x^4 \ln \frac{x^2}{a_{b}}, \nonumber \\
&J_{\text{F}}^{\text{high-T}}(x) = \frac{7 \pi ^4}{360} - \frac{\pi ^2}{24} x^{2} - \frac{1}{32} x^4 \ln \frac{x^{2}}{a_{f}}, \label{2-15}  
\end{align}  
where \( a_b = \pi ^2 \exp \left(\frac{3}{2}-2 \gamma_{E} \right) \) and \( a_f = 16 \, a_b \), with \( \gamma_{E} \) being the Euler-Mascheroni constant. With the logarithmic term included, the high-temperature expansion for the thermal functions is accurate to better than 10 percent even for \( x \sim (1-1.5) \) (depending on the function), but fails beyond that range.
Replacing (\ref{2-15}) in (\ref{2-13}) and disregarding the field-independent term results in
\begin{align}  
&V^{\text{high-T}} = \lambda_4 \phi^4 - \lambda_3 T \phi^3 + \lambda_2 T^2 \phi^2, \label{2-16}  
\end{align}
where
\begin{align}
&\lambda_4 = \frac{1}{64 \pi^2 \nu^4} \left( \sum_{k} g_{k}m_{k}^{4}\ln \frac{a_{b,f} \nu^2}{m_{k}^{2}} \right) + b \ln \frac{T^2}{\nu^2} - b \ln \frac{\phi^2}{\nu^2}, \nonumber \\
& \lambda_3 = \frac{1}{12 \pi \nu^3} \left( \sum_{B} g_{B}m_{B}^{3} \right),\nonumber \\
&\lambda_2 = \frac{1}{24 \nu^2} \left( \sum_{B} g_{B}m_{B}^{2} - \frac{1}{2} \sum_{F} g_{F} m_{F}^{2} \right).
\label{2-17} 
\end{align}

\subsection{Daisy resummation technique}  
Daisy resummation is a technique used to improve the predictions of effective potential calculations, especially at finite temperatures. It aims to sum an infinite series of certain perturbative corrections. The Daisy resummation ensures that the leading order thermal corrections are properly included, addressing the inadequacies of the naive perturbation approach at high temperatures.  

The Daisy resummation term can be articulated as \cite{Carrington:1991hz}
\begin{equation}
V_{D}  = \sum_{B} \frac{\overline{g}_{B} T}{12 \pi} \left(  M_{B}^{3} - (M_{B}^{2}  + \Pi_{B}(T))^{3/2} \right) .  \label{2-18}
\end{equation}
Incorporating \( V_{D} \) involves resumming the infrared-divergent terms associated with the Matsubara zero mode propagator. This is equivalent to replacing \( M^{2} \) with \( M_{tree}^{2} + \Pi (T) \) in the full effective potential, under the assumption that the thermal mass of the zero mode is the only significant factor, which implies utilizing the high-temperature approximation. The summation in Eq.~(\ref{2-18}) considers only the longitudinal degrees of freedom for the gauge bosons and scalars represented by \( \overline{g}_{B} \). This value is equivalent to \( g_{B} \) for bosons, but smaller than \( g_{B} \) for gauge bosons.

We take \( \Pi_{B}(T) = \eta_B^2 T^2 \), where \( \eta_B \)s are determined by the coupling constants of the model. Therefore, the expression for \( V_{D} \) is given by
\begin{equation}
V_{D} = \sum_{B} \frac{\overline{g}_{B} T}{12 \pi} \left( M_{B}^{3} - \eta_B^3 T^3 \left(1+\frac{M_{B}^{2}}{\eta_B^2 T^2}\right)^{3/2} \right).  \label{2-19}
\end{equation}
Next, by using Newton's binomial formula and ignoring the field-independent term, the high-temperature expansion of (\ref{2-19}) becomes
\begin{equation}
V_{D}^{\text{high-T}} = \overline{\lambda}_4 \phi^4 - \overline{\lambda}_3 T \phi^3 + \overline{\lambda}_2 T^2 \phi^2, \label{2-20}  
\end{equation}
where
\begin{align}
\overline{\lambda}_4 = - \frac{1}{32 \pi \nu^4} \left( \sum_{B} \frac{\overline{g}_{B} m_{B}^{4}}{\eta_B} \right), \nonumber \\
\overline{\lambda}_3 = - \frac{1}{12 \pi \nu^3} \left( \sum_{B} \overline{g}_{B} m_{B}^{3} \right), \nonumber \\
\overline{\lambda}_2 = - \frac{1}{8 \pi \nu^2} \left( \sum_{B} \eta_B \overline{g}_{B} m_{B}^{2} \right). \label{2-21}
\end{align}

Finally, combining (\ref{2-3}), (\ref{2-11}), (\ref{2-16}), and (\ref{2-20}) the effective potential at high temperatures takes the form
\begin{equation}
V_{\text{eff}}^{\text{high-T}}(\phi,T) = \frac{1}{4}\lambda(T) \phi^4 - \frac{1}{3} c T \phi^3 + \frac{1}{2} d T^2 \phi^2, \label{2-22}  
\end{equation}
where
\begin{align}
 \lambda(T) &= \frac{1}{16 \pi^2 \nu^4} \left( \sum_{k} g_{k}m_{k}^{4}\ln \frac{a_{b,f} \nu^2}{\sqrt{e}m_{k}^{2}} - \sum_{B} \frac{2 \pi \overline{g}_{B} m_{B}^{4}}{\eta_B}  \right) +4 b \ln \frac{T^2}{\nu^2} \equiv \lambda_{\nu}+4 b \ln \frac{T^2}{\nu^2}, \nonumber \\
 c &=  \frac{1}{4 \pi \nu^3} \left( \sum_{B} (g_{B}-\overline{g}_{B})m_{B}^{3} \right), \nonumber \\
 d &= \frac{1}{12 \nu^2} \left( \sum_{B} (g_{B}-\frac{3 \eta_B \overline{g}_{B}}{\pi})m_{B}^{2} - \frac{1}{2} \sum_{F} g_{F} m_{F}^{2} \right). \label{2-23}
\end{align}
Henceforth, we treat (\ref{2-22}) as the high-temperature effective potential for scale-invariant models, where \( \lambda(T) \), \( c \), and \( d \) are positive dimensionless parameters.

\section{Metastable vacuum decay} \label{sec3}
Metastable vacuum decay refers to the process by which a system transitions from a metastable state—an equilibrium that is stable under small perturbations—to a lower-energy vacuum state. In quantum field theory, vacuums are often characterized by different energy levels, and a metastable vacuum can exist for a considerable amount of time before decaying due to quantum tunneling or thermal fluctuations. This decay can have profound implications in various physical contexts, such as in cosmology, where it could lead to rapid phase transitions affecting the early universe, or in particle physics, where the decay of a metastable vacuum may result in observable phenomena, including the production of massive particles. The timescales for decay can vary dramatically, ranging from extremely short to extraordinarily long, making the study of metastable vacuums crucial for understanding the stability of fundamental theories and the potential for new physics beyond the Standard Model.

\subsection{Analyzing effective potential}
We proceed by introducing dimensionless variables through the substitution of
\begin{equation}
x^{\mu} \to \nu^{-1} x^{\mu}, \quad \phi \to \nu \phi, \quad \text{and} \quad T \to \nu T.
\label{3-1}
\end{equation}
Under these transformations the action changes to,
\begin{equation}
S=\int d^4x \left( \frac{1}{2} \partial_{\mu} \phi\partial^{\mu} \phi - V(\phi,T) \right),
\label{3-2}
\end{equation}
where
\begin{equation}
V(\phi,T)=\frac{1}{4}\lambda_T \phi^4 - \frac{1}{3} c T \phi^3 + \frac{1}{2} d T^2 \phi^2,
\label{3-3}
\end{equation}
with
\begin{equation}
\lambda_T=\lambda_{\nu}+8 b \ln T.
\label{3-4}
\end{equation}
Note that for $ \lambda_T > 0 $, we should have $ T > \exp(-\lambda_{\nu}/8b) $.

To find critical points, i.e., the minima and maxima of the potential (\ref{3-3}), we need to calculate the partial derivative of \( V \) with respect to \( \phi \) and set it to zero,
\begin{equation}
\phi \left( \lambda_T \phi^2 - c T \phi + d T^2 \right) = 0.
\label{3-5}
\end{equation}
This gives us three solutions:  
\begin{equation}
\phi_1 = 0 \quad \text{and} \quad \phi_{2,3} = \frac{c \mp \sqrt{c^2 - 4 \lambda_T d }}{2 \lambda_T}T.
\label{3-6}
\end{equation}
To classify the nature of critical points (whether they are minima or maxima), we calculate the second derivative at critical points 
\begin{equation}
\frac{\partial^2 V}{\partial \phi^2} \bigg|_{\phi_1} = d T^2 \quad \text{and} \quad \frac{\partial^2 V}{\partial \phi^2} \bigg|_{\phi_{2,3}} = \mp \sqrt{c^2 - 4 \lambda_T d } \, T \phi_{2,3}.
\label{3-7}
\end{equation}
Given that \( d > 0 \) and \( T > 0 \), the second derivative at \( \phi_1 \) is positive, which signifies that \( \phi_1 \) represents a local minimum. The nature of the critical points derived from the quadratic equation hinges on the discriminant \( c^2 - 4 \lambda_T d \):  
\begin{itemize}  
    \item If the discriminant is negative, there are no real critical points aside from \( \phi_1 = 0 \).  
    \item If the discriminant is zero, there exists a double root, leading to the condition \( \frac{\partial^2 V}{\partial \phi^2} \bigg|_{\phi_2 = \phi_3} = 0 \).  
    \item If the discriminant is positive, there are three distinct critical points such that \( \phi_1 < \phi_2 < \phi_3 \). In this scenario, according to equation (\ref{3-7}), \( \phi_2 \) acts as a local maximum, while \( \phi_3 \) represents the second local minimum.  
\end{itemize}  
Consequently, the emergence of the second minimum (and a maximum situated between the two local minima) occurs when
\begin{equation}
c^2  - 4 \lambda_T d > 0 \quad \Rightarrow \quad T < \widehat{T} \equiv \exp\left(\frac{c^2  - 4 \lambda_{\nu} d}{32bd}\right),
\label{3-8}
\end{equation}
where we have used (\ref{3-4}). The effective potential at the critical points is
\begin{equation}
V(\phi_1,T) = 0 \quad \text{and} \quad V(\phi_{2,3},T) = \frac{1}{24} \left(c \pm 3 \sqrt{c^2-4 \lambda _T d}\right) T \phi^3_{2,3}.
\label{3-9}
\end{equation}

We now turn our attention to the concept of critical temperature, a key point at which significant changes in a system's behavior occur. The shape of the effective potential \( V(\phi, T) \) is fundamental to understanding the nature of phase transitions.   
At temperatures below \( \widehat{T} \), the effective potential can show two distinct minima, labeled \( \phi_1 \) and \( \phi_3 \), with a maximum in between at \( \phi_2 \). This configuration indicates a first-order phase transition. At the critical temperature \( T_c \), the effective potential values at these two minima become equal, resulting in degeneracy, which can be expressed as
\begin{equation}  
V(\phi_1, T_c) = V(\phi_3, T_c).
\label{3-10}  
\end{equation}  
This condition allows the system to coexist in two distinct states represented by the minima. Taking into account equations (\ref{3-4}), (\ref{3-9}), and (\ref{3-10}), we can derive the expression for the critical temperature as follows
\begin{equation}  
2c^2 - 9 \lambda_{T_c} d = 0 \quad \Rightarrow \quad T_c = \exp\left(\frac{2c^2 - 9 \lambda_{\nu} d}{72bd}\right). 
\label{3-11}  
\end{equation}
According to equations (\ref{3-8}) and (\ref{3-11}), it follows that \( T_c < \widehat{T} \), as expected. At temperatures below \( T_c \), the true vacuum corresponds to \( \phi^3 \neq 0 \), while \( \phi_1 = 0 \) represents a false vacuum.

At the critical temperature, the maximum in the effective potential at \( \phi_2 \) serves as an energy barrier that separates the stable phases. This barrier is critical to the system's dynamics; transitioning between the two states requires the system to gain sufficient energy to overcome this potential maximum. 
In first-order phase transitions, this energy barrier facilitates the emergence of distinct phases. As the temperature fluctuates around the critical temperature \( T_c \), spontaneous fluctuations can nucleate bubbles of one phase within the other. This nucleation process is characteristic of first-order transitions, during which the system transitions abruptly between phases rather than smoothly. The condition for the  phase transition to be strongly first order becomes \cite{Quiros:1999jp}
\begin{equation}  
\frac{\phi_3(T_c)}{T_c}=3\frac{d}{c} \gtrsim 1,
\label{3-12}  
\end{equation}  
which serves as a critical criterion for the occurrence of a strongly first-order phase transition. This condition suggests that the dynamics of the phase transition are significantly influenced by the thermal properties of the scalar fields involved. In various theoretical models, satisfying this condition can lead to the generation of a sufficient amount of baryon asymmetry in the universe, which is essential for understanding the matter-antimatter imbalance observed today.  

\subsection{Decay via instantons}  
Vacuum decay, a phenomenon where a false vacuum state transitions to a lower energy true vacuum, is a crucial area of study in quantum field theory and cosmology. One of the most significant mechanisms proposed for vacuum decay is via instantons, which are non-perturbative solutions of the Euclidean equations of motion that contribute to quantum tunneling processes. These solutions provide a way to calculate the decay rate of the false vacuum, leading to insights into the stability of vacuum states in various theories, including quantum chromodynamics (QCD) and other gauge theories \cite{Coleman:1977py,Callan:1977pt}. The role of instantons in vacuum decay highlights the interplay between quantum effects and classical configurations, offering a framework to understand phenomena such as baryogenesis and the dynamics of early universe phase transitions \cite{Linde:1981zj}.  

We begin with the action for a scalar field defined in a flat spacetime, (\ref{3-2}), where \(V(\phi,T)\) represents a potential function with two local minima. The tunneling process is described by the solution to the Euclidean ($ x_0\rightarrow\tau=it $) equations of motion, given by  
\begin{equation}  
\nabla^2 \phi = \frac{\partial V(\phi,T)}{\partial \phi}.
\label{3-13}  
\end{equation} 
Assuming that the scalar field possesses an \(O(4)\) symmetry, i.e., \(\phi(x) = \phi(r)\) where \(r = \sqrt{\tau^2 + \sum x_i^2}\), the equation simplifies to  
\begin{equation}  
\frac{d^2 \phi}{dr^2} + \frac{D - 1}{r} \frac{d \phi}{dr} = \frac{\partial V(\phi,T)}{\partial \phi},
\label{3-14}  
\end{equation}
where \( D=4 \) represents the number of dimensions in spacetime. This equation describes the motion of a particle that navigates within the "inverted" potential \(-V(\phi,T)\) and is subject to a velocity-dependent "friction" term given by \((D - 1)/r\). The relevant boundary conditions for this system are expressed as  
\begin{equation}  
\phi'(0) = 0, \quad \lim_{r \to \infty} \phi(r) = 0,
\label{3-15}   
\end{equation}
where the prime notation signifies differentiation with respect to \( r \).  
To solve these equations, one typically employs numerical methods, leading to the determination of instanton solutions that characterize the tunneling phenomena. The actions associated with these solutions provide critical insights into the decay rates of the false vacuum states and the stability of the true vacuum configurations. There is an accurate semi-analytical approximation for tunneling action \cite{Adams:1993zs}
\begin{equation}  
S_4= 2 \pi^2 \int_0^{\infty} r^3 dr \left[ \frac{1}{2} \left( \frac{d \phi}{dr} \right)^2 + V(\phi,T) \right]=\frac{4 \pi ^2}{3 \lambda _T} \frac{ \left(\alpha _1 \delta +\alpha _2 \delta^2 + \alpha _3 \delta ^3 \right)}{(2-\delta )^3}
\label{3-16}   
\end{equation}
where $ \delta \equiv \frac{9 \lambda _T d}{c^2} $ with $ \alpha_1 = 13.832 $, $ \alpha_2 = -10.819 $, and $ \alpha_3 = 2.0765 $. Note that the tunneling process begins when $ T < T_c $, or equivalently $ \lambda_T < \lambda_{T_c} $, therefore according to (\ref{3-11}) $ 0 < \delta < 2 $. Finally, the decay rate per unit volume of the false vacuum through instantons is
\begin{equation}  
\Gamma_{\text{ins}}(T) \simeq R_0^{-4} \left( \frac{S_4}{2\pi} \right)^{2} \exp(-S_4),
\label{3-17}   
\end{equation}
where \( R_0 \simeq \mathcal{O}(\phi_3^{-1}) \) signifies the size of the bubble.

\subsection{Decay via sphalerons}  
Sphalerons constitute a significant category of non-perturbative solutions in quantum field theory, especially critical in the study of baryon number violation and electroweak interactions. In contrast to instantons, which facilitate tunneling processes between different vacuum states, sphalerons correspond to static configurations that exist at the peaks of energy barriers separating local minima in the potential landscape. These solutions are pivotal for understanding mechanisms such as baryogenesis, which involves the generation of baryon asymmetry in the early universe, as they allow for transitions that can change the baryon number of a system. The stability of sphalerons is intimately linked to thermal fluctuations, making them essential in analyzing phase transitions in the early universe, where temperatures were high enough to allow such transitions to occur. 

In the context of a scalar field theory with a potential that has multiple minima, sphalerons can be viewed as unstable equilibrium points in the potential landscape. The action for sphalerons can be derived similarly to instantons, but instead of considering a tunneling process, we analyze the system's behavior at the top of the potential barrier.  

To describe sphalerons, we can start from the action defined in the previous subsection with the potential \( V(\phi,T) \). The equations of motion remain the same as in (\ref{3-14}) with $ D=3 $. For sphalerons, we look for solutions that correspond to static configurations in Euclidean time. Specifically, we consider configurations where the scalar field approaches the local minima of the potential at spatial infinity.
    
The sphaleron solutions can be characterized by their energy, which is related to the height of the potential barrier. The action associated with a sphaleron configuration can be computed similarly to that of instantons, but it is typically larger due to the static nature of the solutions. A precise semi-analytical approximation exists for the sphaleron action \cite{Adams:1993zs},

\begin{equation}  
\frac{S_3}{T}= \frac{4 \pi}{T} \int_0^{\infty} r^2 dr \left[ \frac{1}{2} \left( \frac{d \phi}{dr} \right)^2 + V(\phi,T) \right]=\frac{64 \pi \sqrt{d}}{81 \lambda _T} \frac{\left(\beta _1 \delta +\beta _2 \delta^2 + \beta _3 \delta ^3 \right)}{(2-\delta )^2},
\label{3-18}  
\end{equation}  
where $ \beta_1 = 8.2938, $, $ \beta_2 = -5.5330 $, and $ \beta_3 = 0. 8180 $. 

The rate of sphaleron transitions can be expressed in terms of the sphaleron action, leading to a decay rate analogous to that of instantons. However, the decay mechanism differs fundamentally, as it does not involve tunneling through a potential barrier but rather transitions over the barrier facilitated by thermal fluctuations. The rate per unit volume can be approximated as \cite{Linde:1981zj}
\begin{equation}  
\Gamma_{\text{sph}}(T) \simeq T^4 \left( \frac{S_3}{2\pi T} \right)^{\frac{3}{2}} \exp\left(-\frac{S_3}{T}\right) .
\label{3-19}  
\end{equation}
The exponential term captures the suppression of the decay rate at lower temperatures due to the Boltzmann factor. As \(S_3\) increases, the rate \(\Gamma_{\text{sph}}(T)\) decreases exponentially for a fixed temperature, implying that larger action corresponds to a lower likelihood of decay.

\begin{figure}[H]
\centerline{\hspace{0cm}\epsfig{figure=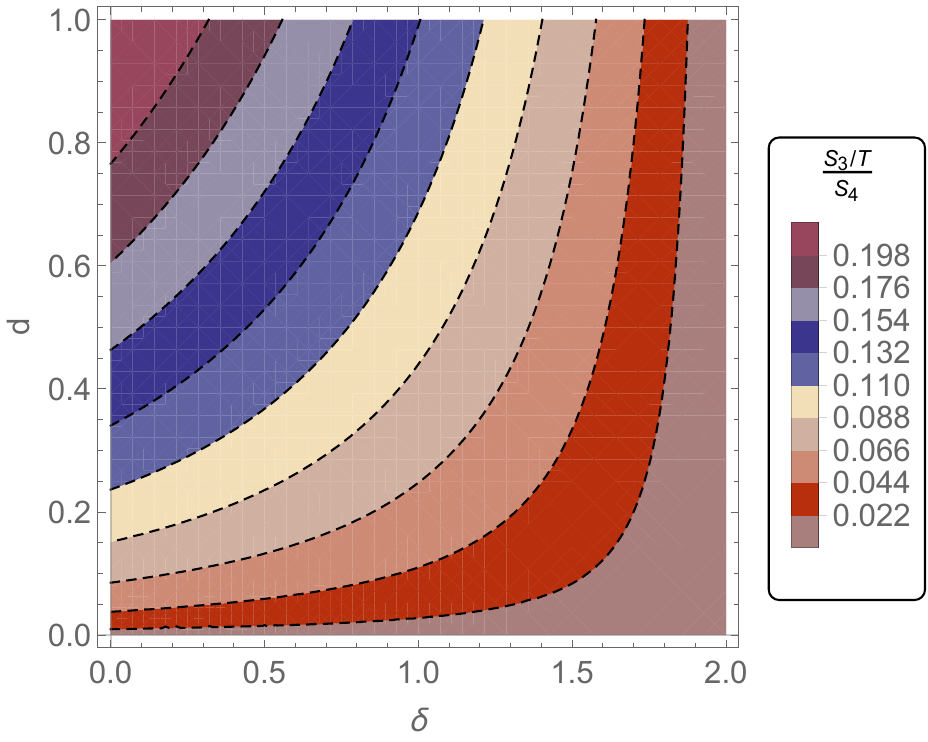,width=7cm}}
\caption{The ratio \((S_3/T)/S_4\) plotted in the \(\delta\)-\(d\) plane, indicating that quantum tunneling is less significant than sphaleron processes, as reflected by \((S_3/T) < S_4\) in all regions. This supports the use of equation (\ref{3-19}) for estimating the vacuum decay rate.
} \label{fig1}
\end{figure}

\begin{figure} [H] 
\centerline{\hspace{0cm}\epsfig{figure=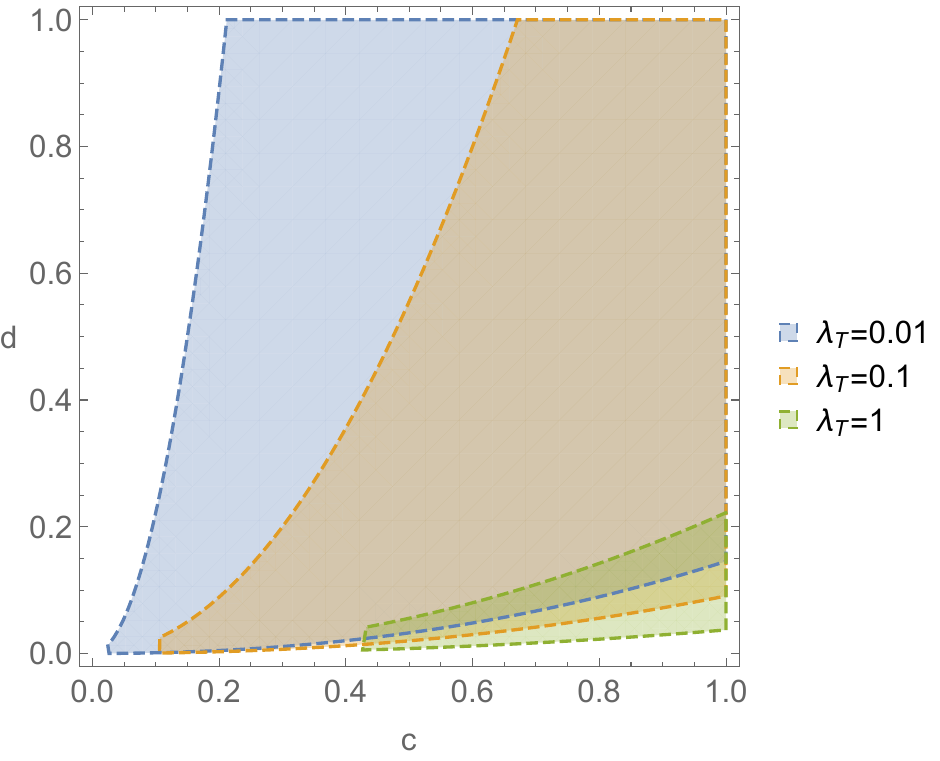,width=7cm}}
\caption{Regions where the decay rate via sphalerons exceeds 100 times that of instantons for three different \( \lambda_T \). In nearly all areas where \( 0 < \delta < 2 \), sphalerons dominate over instantons.
} \label{fig2}
\end{figure}

Figure \ref{fig1} shows the ratio \((S_3/T)/S_4\) in the \(\delta-d\) plane, demonstrating that \((S_3/T) < S_4\) in all regions. This indicates that quantum tunneling (\ref{3-17}) is less significant, and the vacuum decay rate should be estimated using (\ref{3-19}). Figure \ref{fig2} illustrates the parameter space \(c\) and \(d\) where the decay rate through sphalerons exceeds 100 times that of instantons for three different values of \(\lambda_T\). Sphalerons dominate the decay process in nearly all regions defined by \(0 < \delta < 2\), underscoring their significant role in high-temperature phase transitions.

\subsection{Phase transition dynamics}
In the context of first-order phase transitions, the dynamics involve overcoming a potential barrier that separates distinct minima due to thermal effects. This process is characterized by the nucleation of a broken phase within an unstable phase. The critical aspect of such transitions is determining the time of nucleation, at which the likelihood of forming a true vacuum bubble within a horizon radius becomes significant. This is quantitatively expressed by the integral \cite{Ellis:2018mja}
\begin{equation}  
N(T_n) = \int_{T_n}^{T_c} \frac{dT}{T} \frac{\Gamma(T)}{H(T)^4} = 1,
\label{3-20}  
\end{equation}
where \(T_n\) represents the nucleation temperature. This relationship highlights the interplay between the decay rate \(\Gamma(T)\) given in (\ref{3-19}) and the Hubble parameter \(H(T)\), essential for understanding the emergence of stable bubbles in the thermal landscape of the early universe. 

The Hubble expansion rate \(H(T)\) during the radiation-dominated epoch plays a crucial role in the dynamics of first-order phase transitions. It can be expressed as 
\begin{equation}  
H(T)^2 = \frac{\rho_{\text{rad}}}{3M_P^2},
\label{3-21}  
\end{equation}
where \(\rho_{\text{rad}}\) signifies the radiation energy density of relativistic particle species, defined as 
\begin{equation}  
\rho_{\text{rad}} \equiv \frac{\pi^2}{30} g_* T^4.
\label{3-22}  
\end{equation}
Here, \(M_P = 2.4 \times 10^{18}\) GeV denotes the reduced Planck mass, and \( g_* \simeq \mathcal{O}(100) \) accounts for the temperature-dependent degrees of freedom contributing to the effective energy density.

\begin{figure} [H] 
\centerline{\hspace{0cm}\epsfig{figure=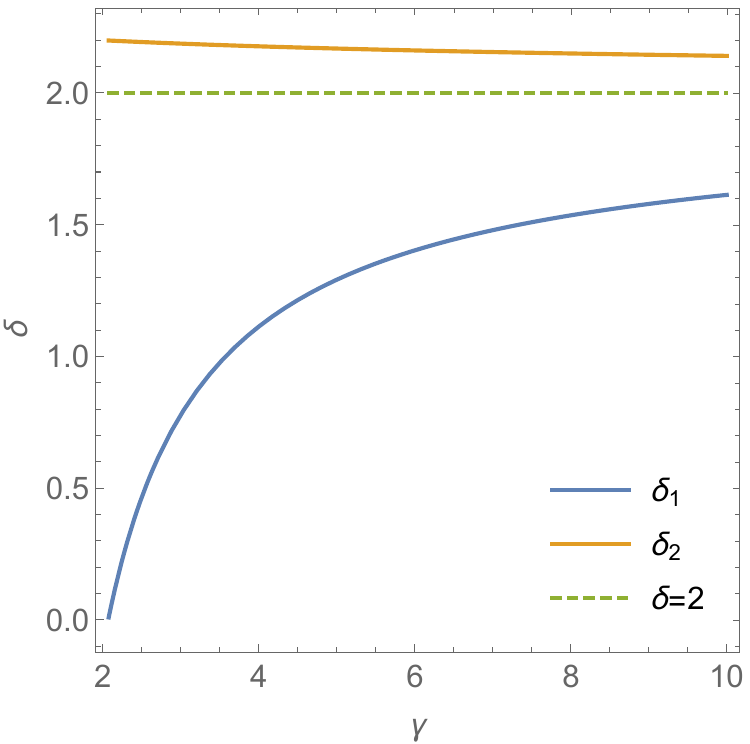,width=7cm}}
\caption{Representation of the relationship between the parameters \( \delta_1 \) and \( \delta_2 \) under varying values of \( \gamma \). In both cases $ \displaystyle{\lim_{\gamma \to \infty}} \delta_{1,2} = 2$.}
\label{fig3}
\end{figure}

\begin{figure} [H] 
\centerline{\hspace{0cm}\epsfig{figure=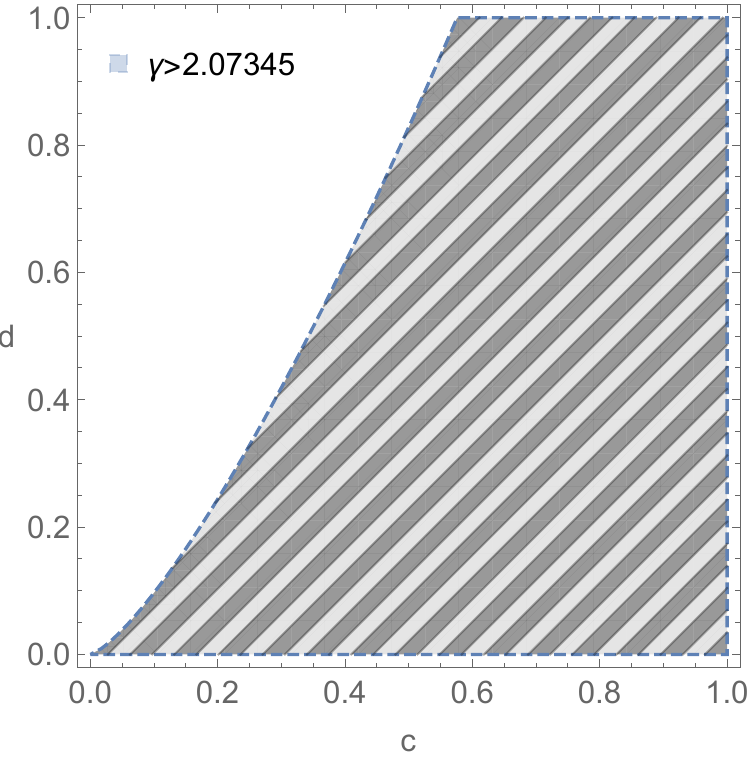,width=5.7cm}\hspace{0cm}\epsfig{figure=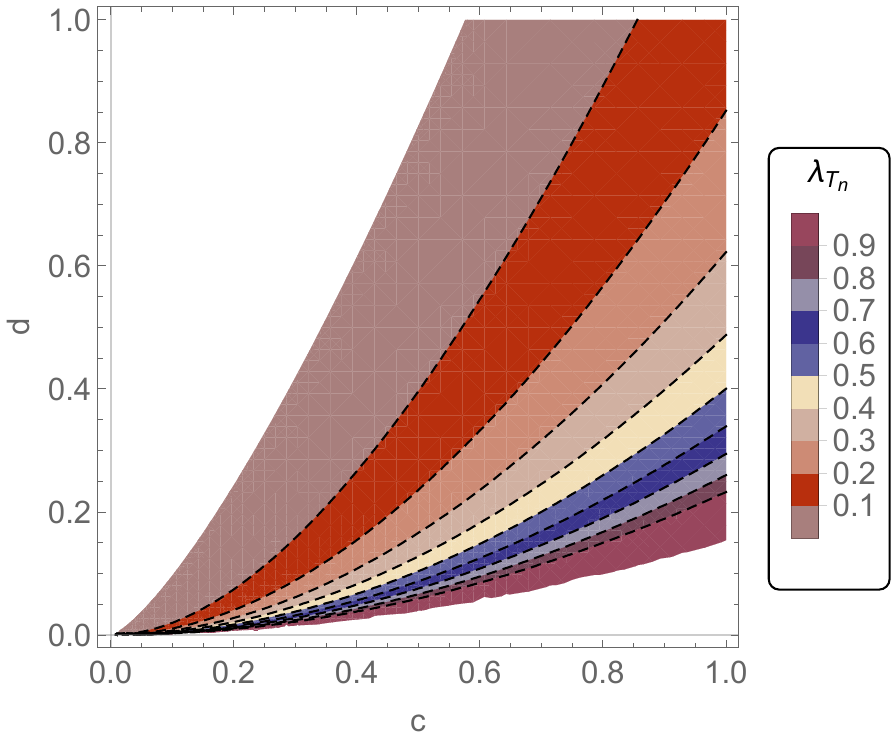,width=7cm}}
\caption{Left: visualization of the allowable parameter space for \( c \) and \( d \) constrained by the condition \( \gamma > 2.07345 \). The shaded regions represent the combinations of \( c \) and \( d \) that satisfy the given criteria, allowing for a better understanding of how different parameter values influence the system's behavior in conjunction with the inequalities associated with \( \delta_1 \) and \( \delta_2 \). Right: $ \lambda_{T_n} $ as a function of the parameter space $ c $ and $ d $.}
\label{fig4}
\end{figure}

The study of nucleation temperature, \( T_n \) defined in (\ref{3-20}), has seen significant advancements, particularly through various approximations that simplify calculations by bypassing complex integrations \cite{Athron:2022mmm}. Notably, the most basic approximation relates the bounce action to the nucleation temperature through the formula \cite{Quiros:1999jp}
\begin{equation}
\frac{S_3(T_n)}{T_n} \sim 140.
\label{3-23}  
\end{equation}
This condition, derived from the requirement that the bubble nucleation rate \( \Gamma(T) \sim T^4 e^{-S_3(T)/T} \) becomes efficient (i.e., \( \Gamma \sim H^4 \)), provides a simple and widely adopted criterion for estimating \( T_n \).
Substituting (\ref{3-18}) in (\ref{3-23}) gives
\begin{equation}
\delta_{1,2} = \frac{\beta _2+4 \gamma \mp \sqrt{8 \beta _2 \gamma +4 \beta _1 \left(\gamma -\beta _3\right)+16 \beta _3 \gamma +\beta _2^2} }{2 \left(\gamma -\beta _3\right)}
\label{3-24}  
\end{equation}
where \( \gamma \equiv 140 \frac{9 c^2}{64 \pi d^{3/2}} \). The constraint \( 0 < \delta < 2 \) is not satisfied by \( \delta_{2} \), indicating its inapplicability under the specified conditions. In contrast, \( \delta_{1} \) adheres to the inequality when \( \gamma > 2.07345 \); for a visual representation, refer to figure \ref{fig3}. This threshold condition of \( \gamma > 2.07345 \) imposes limitations on the parameter space defined by \( c \) and \( d \), narrowing the feasible combinations of these variables. The allowable regions of this parameter space as well as
\begin{equation*}
\lambda_{T_n} = \frac{c^2 \delta_1}{9d},
\end{equation*}
are illustrated in figure \ref{fig4}, providing further insight into the relationships among these parameters and their impact on the system.

\begin{figure} [H] 
\centerline{\hspace{0cm}\epsfig{figure=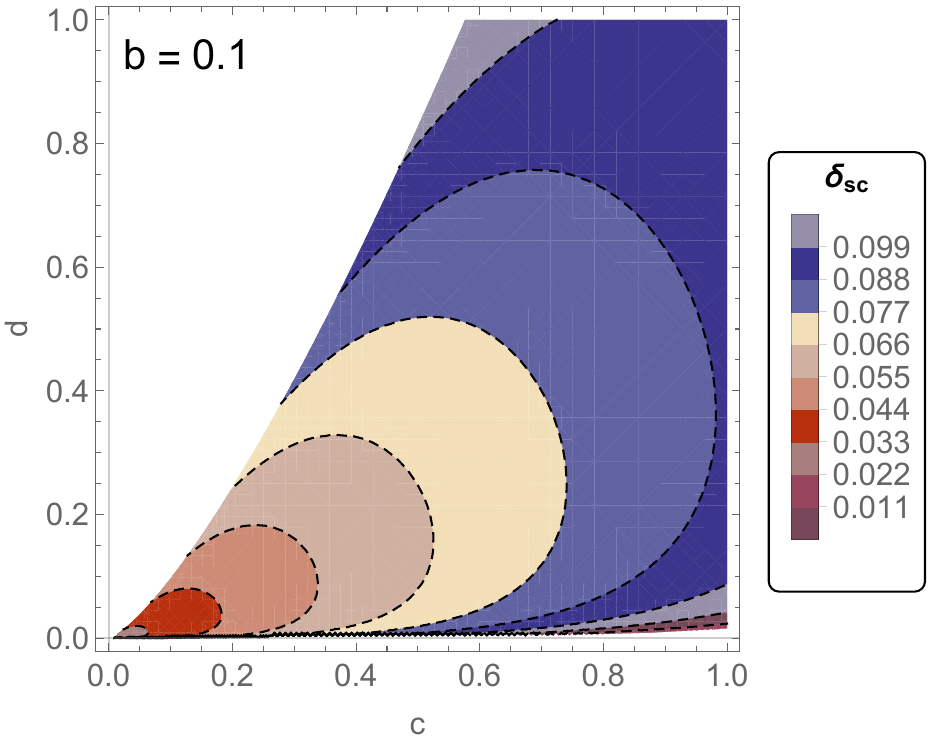,width=6.5cm}\hspace{0cm}\epsfig{figure=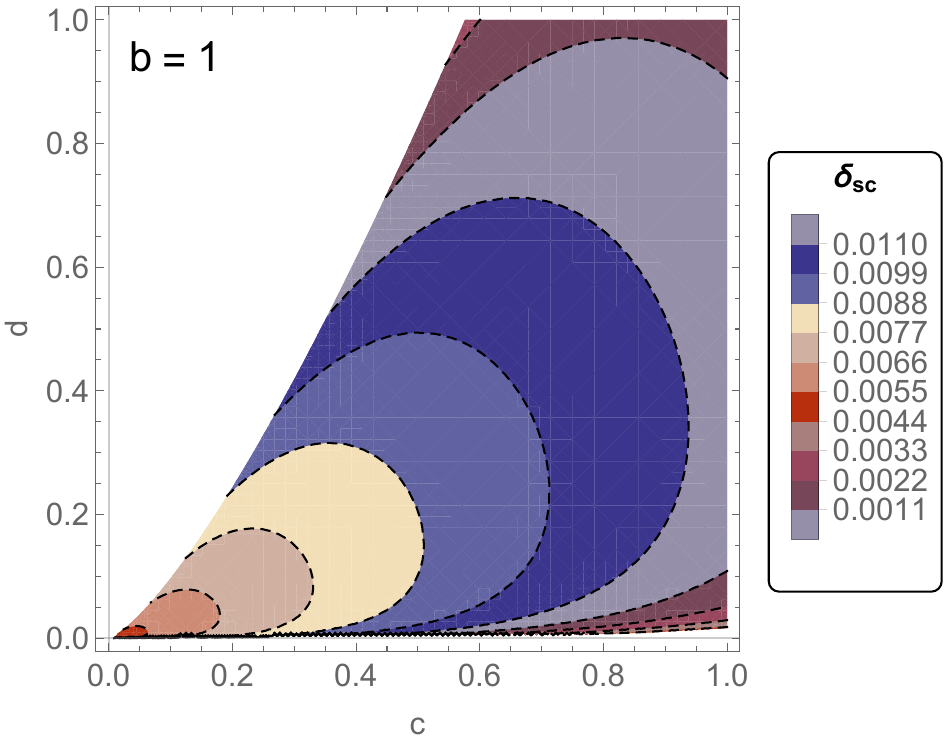,width=6.5cm}}
\caption{Illustration of the relationship between the critical temperature \( T_c \) and the nucleation temperature \( T_n \). This figure presents how variations in the parameters \( b \), \( c \), and \( d \) influence the supercooling  parameter \( \delta_{\text{sc}} \).}
\label{fig5}
\end{figure}

As the universe expands and cools, the temperature can drop significantly below the critical temperature \( T_c \) without the system transitioning to a different phase. This process, known as supercooling, can lead to a variety of physical implications.
Using (\ref{3-4}), (\ref{3-11}), and \( \lambda_{T_n} = \frac{c^2 \delta_1}{9d} \), the supercooling parameter \cite{Athron:2022mmm} is given by 
\begin{equation}
\delta_{\text{sc}} \equiv \frac{T_c - T_n}{T_c}=1-\exp \left[  -\frac{c^2}{72bd}(2-\delta_1)\right] .
\label{3-25}  
\end{equation}
This equation encapsulates the relationship between the critical temperature \( T_c \), the nucleation temperature \( T_n \), and three other parameters, $ b $, $ c $, and $ d $ that characterize the dynamics of the transition, see figure \ref{fig5}.

In analyzing vacuum energy, also known as latent heat density, we can utilize the conventional statistical mechanics expression given by  
\begin{equation}  
\rho_{\text{vac}} = -V(\phi_3, T) + \frac{1}{4} T \frac{d}{dT} V(\phi_3, T) = \frac{81 b d^4 \left(9-2 \delta +3 \sqrt{9-4 \delta }\right)^2 T^4}{8 c^4 \delta ^4}  ,
\label{3-26}  
\end{equation}  
where the equality in the latter part is derived from equations (\ref{3-4}), (\ref{3-6}), and (\ref{3-9}), with the substitution \( \lambda_T \to \frac{c^2 \delta}{9 d} \). This expression for \(\rho_{\text{vac}}\) encapsulates the thermodynamic characteristics of the vacuum energy process effectively. To better understand the implications of this energy density, we can establish a ratio between the vacuum energy density and the radiation energy density at the nucleation temperature. This ratio, denoted as \( \alpha \), is expressed as
\begin{equation}
\alpha \equiv \frac{\rho_{\text{vac}}}{\rho_{\text{rad}}} \bigg|_{T=T_n} =  \frac{1215 b d^4 \left(9-2 \delta_1 +3 \sqrt{9-4 \delta_1 }\right)^2}{4 \pi^2 g_* c^4 \delta_1 ^4},
\label{3-27}  
\end{equation}
providing a quantitative measure of the relationship between these two forms of energy in the context of phase transitions and thermal dynamics. This fundamental relationship is crucial for understanding the energy distribution and conversion processes that occur during such transitions.

Moreover, the bounce action provides insights into estimating the duration of the phase transition. The inverse duration in Hubble units is pivotal for analyzing the transition dynamics. This ratio is typically derived by performing a Taylor expansion of the exponent in eq.~(\ref{3-19}) around a specific time, $t_*$. 
The transition rate $\Gamma$ can be expressed as
\[
\Gamma \propto e^{-\frac{S_3}{T}} = e^{\beta_0 + \beta_* H(T_*) (t - t_*) + \ldots}.
\]
By differentiating both sides with respect to time $t$, we can derive an expression for $\beta_* $
\begin{equation}
\beta_* = - \frac{1}{H(T)} \frac{d}{dt}\left(\frac{S_3}{T}\right)  \bigg|_{T=T_*}=  T \frac{d}{dT}\left(\frac{S_3}{T}\right)  \bigg|_{T=T_*}.
\label{3-28}  
\end{equation}
This equation highlights the critical role of \( \beta_* \) in describing the dynamics of phase transitions and the significance of understanding its behaviors in the context of supercooling and nucleation phenomena. At the nucleation temperature we define
\begin{equation}
\beta \equiv  T \frac{d}{dT}\left(\frac{S_3}{T}\right) \bigg|_{T=T_n} = \frac{512 \pi  b d^{5/2} \left(2 \beta _1 +  (2+\delta_1)\beta _2 +4  \delta_1 \beta _3 \right)}{c^4 (2 - \delta_1)^3},
\label{3-29}  
\end{equation}
where the equality in the latter portion arises from equations (\ref{3-4}) and (\ref{3-18}), using the substitution \( \lambda_T \to \frac{c^2 \delta}{9 d} \). According to (\ref{3-29}) $ \beta $ is positive for $ 0<\delta_1<2 $.

\begin{figure} [H] 
\centerline{\hspace{0cm}\epsfig{figure=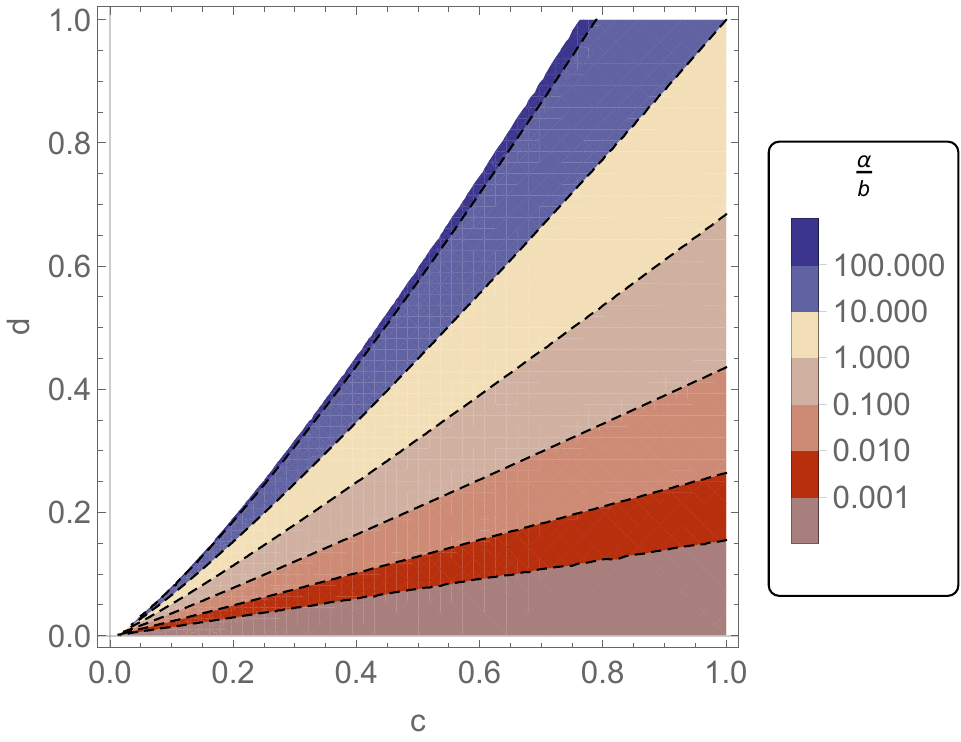,width=6.5cm}\hspace{0cm}\epsfig{figure=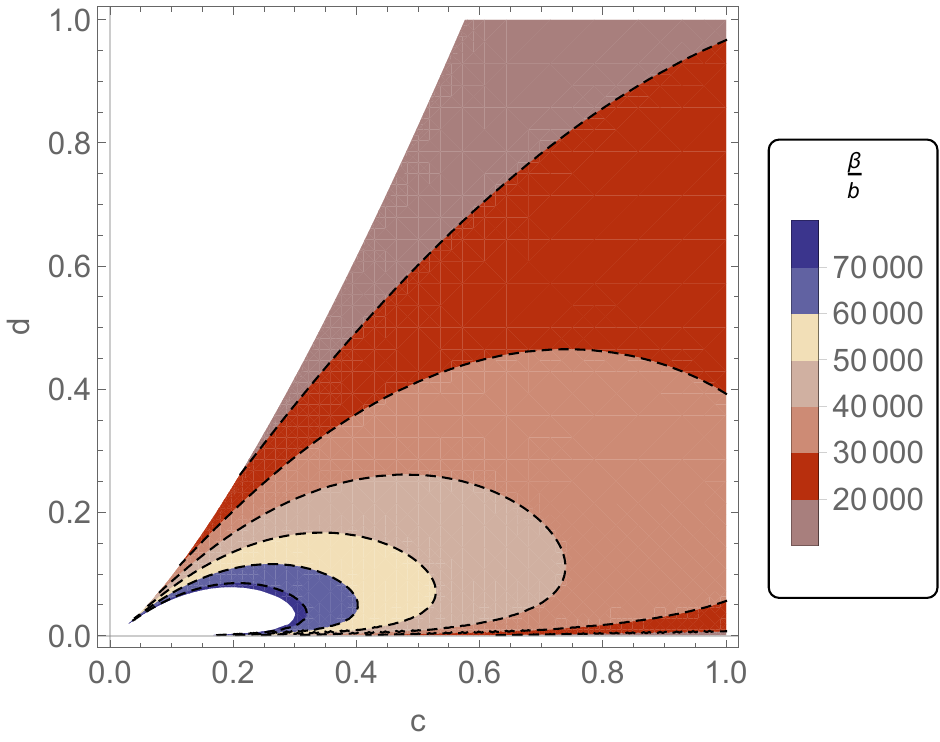,width=6.35cm}}
\caption{Variation of the parameters \( \alpha \) and \( \beta \) as functions of the model parameters. The left panel shows the contour plot of \( \alpha/b \), and the right panel depicts the behavior of \( \beta/b \).}
\label{fig6}
\end{figure}

The parameters \( \alpha \) and \( \beta \) play crucial roles in understanding the dynamics of phase transitions and their implications for GW (GW) production in the early universe. The ratio \( \alpha \), which quantifies the relationship between vacuum energy density and radiation energy density at the nucleation temperature, is essential for assessing the energy landscape during phase transitions. A high \( \alpha \) value indicates a significant contribution of vacuum energy, which can influence the dynamics of bubble nucleation and expansion, ultimately affecting the GW spectrum generated during the transition. On the other hand, \( \beta \) characterizes the rate of change of the bounce action with respect to temperature, providing insights into the duration of the phase transition. A larger \( \beta \) signifies a more rapid transition, which is pivotal for the generation of GWs. Figure \ref{fig6} illustrates how both \( \alpha \) and \( \beta \) vary as functions of the parameters \( b \), \( c \) and \( d \).

In the next section we discuss the impact of \( \alpha \) and \( \beta \) on the GW signals that may be detectable in future experiments.

\section{Gravitational Waves} \label{sec4}

GWs generated during a first-order phase transition arise from multiple processes, including bubble collisions, sound waves, and turbulence. Each of these mechanisms contributes to the stochastic GW background, and their combined spectrum can potentially be observed by future space-based detectors.

In the study of cosmological phase transitions, the nucleation temperature \( T_n \) is often used as a proxy for the temperature \( T_* \) at which GWs are produced, particularly in scenarios without significant reheating. This approximation, \( T_* \approx T_n \), is justified for transitions occurring in a radiation-dominated epoch, where the time between bubble nucleation and percolation is short, and reheating effects are negligible.  While the percolation temperature \( T_p \) more accurately marks the onset of bubble collisions and GW production, the nucleation temperature serves as a reasonable approximation in many cases, especially for strong transitions, see eq.~(\ref{3-12}), where \( T_n \approx T_p \). This simplification remains practical for phenomenological studies when detailed percolation dynamics are not required.

Below, we discuss the contributions from bubble collisions, turbulence, and sound waves, and present the combined GW spectrum.

\subsection{Bubble collisions}

When bubbles of the new phase nucleate and expand during a first-order phase transition, their collisions produce significant anisotropic energy-momentum fluctuations, leading to GWs. The GW spectrum resulting from bubble collisions is given by \cite{Huber:2008hg}
\begin{equation}
h^2 \Omega_{\text{coll}}(f) = 1.67 \times 10^{-5} \beta^{-2} \left( \frac{\kappa \alpha}{1 + \alpha} \right)^2  \left( \frac{g_*}{100} \right)^{-\frac{1}{3}} \left( \frac{0.11 \, v_w^3}{0.42 + v_w^2} \right) S_{\text{coll}},
\label{4-1}  
\end{equation}
where \( S_{\text{coll}} \) describes the spectral shape of the GW signal from bubble collisions
\begin{equation}
S_{\text{coll}} = \frac{3.8 \, (f/f_{\text{coll}})^{2.8}}{2.8 \, (f/f_{\text{coll}})^{3.8} + 1}.
\label{4-2}
\end{equation}
with
\begin{equation}
f_{\text{coll}} = 1.65 \times 10^{-5} \beta \left( \frac{0.62}{v_w^2 - 0.1 v_w + 1.8} \right) \left( \frac{T_n}{100} \right) \left( \frac{g_\ast}{100} \right)^{1/6} \, \text{Hz}.
\label{4-3}
\end{equation}
This formula represents the peak frequency of the GW spectrum from bubble collisions during a first-order phase transition. It depends on the bubble wall velocity \( v_w \), the inverse time duration of the phase transition in Hubble units \( \beta \), the nucleation temperature \( T_n \), and the effective number of relativistic degrees of freedom \( g_\ast \).

\subsection{Turbulence contribution}

Turbulence in the plasma, which forms after bubble collisions, also generates GWs. Although this contribution is generally smaller than that from sound waves, it can still be significant. The GW spectrum from turbulence is expressed as \cite{Caprini:2009yp}:

\begin{equation}
h^2 \Omega_{\text{turb}}(f) = 3.35 \times 10^{-4} \beta^{-1} \left( \frac{\kappa_{\text{turb}} \alpha}{1 + \alpha} \right)^{3/2} \left( \frac{g_*}{100} \right)^{-\frac{1}{3}} v_w \, S_{\text{turb}},
\label{4-4}
\end{equation}
where \( S_{\text{turb}} \) characterizes the spectral shape of the turbulence contribution:

\begin{equation}
S_{\text{turb}} = \frac{(f/f_{\text{turb}})^3}{(1 + 8\pi f / h_*)(1 + f / f_{\text{turb}})^{11/3}},
\label{4-5}
\end{equation}
with
\begin{equation}
f_{\text{turb}} = 2.27 \times 10^{-5} \beta \frac{1}{v_w} \left( \frac{T_n}{100} \right) \left( \frac{g_*}{100} \right)^{1/6} \, \text{Hz},
\label{4-6}
\end{equation}
and
\begin{equation}
h_* = 1.65 \times 10^{-5} \left( \frac{T_n}{100} \right) \left( \frac{g_*}{100} \right)^{1/6},
\label{4-7}
\end{equation}
which represents the inverse of the Hubble time at the moment of GW production, redshifted to the present time.

\subsection{Sound waves}

Sound waves in the plasma, generated by the bulk motion of the fluid following bubble collisions, are the primary source of GWs in some scenarios. The GW spectrum from sound waves is given by \cite{Hindmarsh:2015qta}:

\begin{equation}
h^2 \Omega_{\text{sw}}(f) = 2.65 \times 10^{-6} \beta^{-1} \left( \frac{\kappa_v \alpha}{1 + \alpha} \right)^2  \left( \frac{g_*}{100} \right)^{-\frac{1}{3}} v_w \, S_{\text{sw}},
\label{4-8}
\end{equation}
where \( S_{\text{sw}} \) describes the spectral shape of the sound wave contribution:
\begin{equation}
S_{\text{sw}} = (f/f_{\text{sw}})^3 \left( \frac{7}{3 \, (f/f_{\text{sw}})^2 + 4} \right)^{3.5}.
\label{4-9}
\end{equation}
The peak frequency \( f_{\text{sw}} \) is determined by 

\begin{equation}
f_{\text{sw}} = 1.9 \times 10^{-5} \beta \frac{1}{v_w} \left( \frac{T_n}{100} \right) \left( \frac{g_*}{100} \right)^{1/6} \, \text{Hz},
\label{4-10}
\end{equation}
depending on the nucleation temperature \( T_n \), the inverse time duration of the phase transition \( \beta \), and the bubble wall velocity \( v_w \).

\subsection{Combined spectrum}

The total GW spectrum is the sum of the contributions from bubble collisions, sound waves, and turbulence:

\begin{equation}
h^2 \Omega_{\text{tot}} \simeq h^2 \Omega_{\text{coll}} + h^2 \Omega_{\text{sw}} + h^2 \Omega_{\text{turb}}.
\label{4-11}
\end{equation}
where $\kappa$, $\kappa_v$, and $\kappa_{\text{turb}}$,
\begin{align}
\kappa &= \frac{1}{1 + 0.715\,\alpha} \left( 0.715\,\alpha + \frac{4}{27} \sqrt{ \frac{3\alpha}{2} } \right), \nonumber \\
\kappa_v &= \frac{\alpha}{0.73 + 0.083 \sqrt{\alpha} + \alpha}, \quad 
\kappa_{\text{turb}} = 0.05\,\kappa_v,
\end{align}
represent the portions of the latent heat that are converted into: the gradient energy of the Higgs-like field, the kinetic energy of the fluid (bulk motion), and magnetohydrodynamic (MHD) turbulence, respectively.

We proceed to analyze the dependence of GWs on key parameters. The parameter $\lambda_{T_n}$ is fully determined once the parameters $c$ and $d$ are fixed. These parameters govern the dynamics of the phase transition and the shape of the effective potential. However, the parameters $\alpha$ and $\beta$—which quantify the strength of the phase transition and its duration, respectively—depend not only on $c$ and $d$ but also on the parameter $b$. The parameter $b$ plays a crucial role in setting the energy scale of the phase transition and influences the thermal behavior of the scalar field.

A significant constraint arises when attempting to set the nucleation temperature $T_n$ solely through the parameters $b$, $c$, and $d$. This is not possible, as clearly illustrated by the expression:
\begin{equation}
T_n = \nu \exp \left( \frac{\lambda_{T_n} - \lambda_{\nu}}{8b} \right),
\label{4-13}
\end{equation}
which shows that $T_n$ depends explicitly on the prefactor $\nu$ and $\lambda_{\nu}$, in addition to $b$ and $\lambda_{T_n}$.  Thus, to uniquely determine $T_n$, one must also have knowledge of $\nu$ and $\lambda_{\nu}$, which are not fixed by $b$, $c$, and $d$ alone.

Since our approach is model-independent, we treat the parameters $b$, $c$, and $d$ as independent quantities. In any specific particle physics model, these parameters would be fixed by fundamental quantities like masses and coupling constants. We further assume that the nucleation temperature $T_n$ can be determined independently of $b$, $c$, and $d$.

For our numerical analysis, we adopt the following benchmark values, which are characteristic of a strong first-order electroweak phase transition:
\begin{equation}
T_n \simeq 100 \, \text{GeV}, \quad g_* \simeq 100, \quad v_w \simeq 1,
\label{4-14}
\end{equation}
where:
\begin{itemize}
    \item[$T_n$:] is the nucleation temperature,
    \item[$g_*$:] is the number of relativistic degrees of freedom,
    \item[$v_w$:] is the wall velocity of the expanding bubble.
\end{itemize}
This choice is well-motivated: a nucleation temperature $T_n \sim 100$\,GeV is natural for electroweak-scale phenomena, $g_* \simeq 100$ corresponds to the Standard Model's particle content at high temperatures, and $v_w \simeq 1$ (the speed of light) indicates a highly relativistic bubble wall, typical of strongly supercooled phase transitions.

To comprehensively explore the parameter space, we conduct a numerical analysis for two distinct values of the parameter $b$, namely $b = 10^{-7}$ and $b = 10^{-2}$. These values are chosen to represent markedly different physical regimes. For each case, we generate a large ensemble of random $(c, d)$ pairs, uniformly sampled from the domain $[0.01, 1] \times [0.01, 1]$. Each candidate point is subjected to a physical viability check, requiring $\lambda_{T_{n}} < 1$ and $\gamma > 2.07345$ as illustrated in Figure~\ref{fig4}. Points failing these criteria are discarded. For each surviving parameter set, we calculate the subsequent cascade of relevant quantities, including the transition strength $\alpha$, the inverse duration $\beta$, the peak GW frequencies ($f_{\text{coll,max}}$, $f_{\text{sw,max}}$, $f_{\text{turb,max}}$), and their corresponding energy densities ($\Omega_{\text{coll,max}}$, $\Omega_{\text{sw,max}}$, $\Omega_{\text{turb,max}}$). The extreme minimum and maximum values observed for each parameter across the entire scanned region are reported in Table~\ref{table}, delineating the theoretically allowed ranges for each scenario.

\setlength{\arrayrulewidth}{0.3pt}
\arrayrulecolor{blue!60!black}
\begin{table}[htbp] 
\centering 
\caption{Extreme values of GW signatures across the viable parameter space. The table contrasts the dramatically different ranges of the peak frequencies $(f)$ and energy densities $(\Omega)$ for the collision, sound wave, and turbulence contributions, generated by the two distinct choices of the parameter $b$.} 
\label{table} 
\vspace{0.3cm}
\begin{NiceTabular}{>{\centering\arraybackslash}m{2.5cm}  
    S[table-format=1.3e-2]
    S[table-format=1.3e-1]
    S[table-format=1.3e-2]
    S[table-format=1.3e-1]
}[code-before = 
    \rowcolor{blue!8}{1}
    \rowcolor{blue!8}{2}
    \rowcolor{gray!10}{3,5,7,9,11,13,15}
    \rowcolor{gray!3}{4,6,8,10,12,14}
]
\toprule
\Block{2-1}{\textbf{Parameter}} & 
\multicolumn{2}{c}{\textbf{$b = 10^{-7}$}} & 
\multicolumn{2}{c}{\textbf{$b = 10^{-2}$}} \\
\cmidrule(lr){2-3} \cmidrule(lr){4-5}
 & \textbf{Min} & \textbf{Max} & \textbf{Min} & \textbf{Max} \\
\midrule
$c$                      & 4.575e-2 & 9.996e-1 & 3.490e-2 & 9.995e-1 \\
$d$                      & 1.991e-2 & 9.975e-1 & 1.131e-2 & 9.982e-1 \\
$\lambda_{T_{n}}$        & 4.094e-4 & 9.964e-1 & 1.206e-4 & 9.991e-1 \\
$\alpha$                 & 4.111e-12 & 6.888e1  & 2.151e-7 & 5.139e9  \\
$\beta$                  & 1.070e-3 & 1.409e-2 & 1.045e2  & 1.520e3  \\
$f_{\text{coll,max}}$    & 4.053e-9 & 5.339e-8 & 3.960e-4 & 5.760e-3 \\
$f_{\text{sw,max}}$      & 2.033e-8 & 2.678e-7 & 1.986e-3 & 2.888e-2 \\
$f_{\text{turb,max}}$    & 8.899e-8 & 6.785e-7 & 2.848e-3 & 4.141e-2 \\
$\Omega_{\text{coll,max}}$ & 5.171e-38 & 1.109e0  & 4.346e-34 & 1.185e-10 \\
$\Omega_{\text{sw,max}}$ & 1.878e-49 & 2.292e-3 & 1.076e-35 & 2.535e-8  \\
$\Omega_{\text{turb,max}}$ & 5.272e-39 & 5.045e-4 & 1.412e-34 & 7.923e-13 \\
$f_{\text{max}}$         & 4.054e-9 & 5.379e-8 & 1.531e-3 & 2.888e-2 \\
$\Omega_{\text{tot,max}}$ & 5.185e-38 & 1.110e0  & 4.639e-34 & 2.538e-8  \\
\bottomrule
\end{NiceTabular}
\end{table}

We consistently observe the following hierarchical relationship among the peak frequencies of the GW spectra generated by different sources:
\[
f_{\text{coll,max}} < f_{\text{sw,max}} < f_{\text{turb,max}},
\]
which holds across the parameter space. This ordering is physically expected because bubble collisions typically produce lower-frequency signals due to the longer duration of the collision process, while sound waves (sw) and magnetohydrodynamic turbulence (turb) involve shorter dynamical timescales and hence higher frequencies. This hierarchy is an important observational signature that can help discriminate between different GW sources in future detectors.

\begin{figure}[H]
    \centerline{\hspace{0cm}\includegraphics[width=13cm]{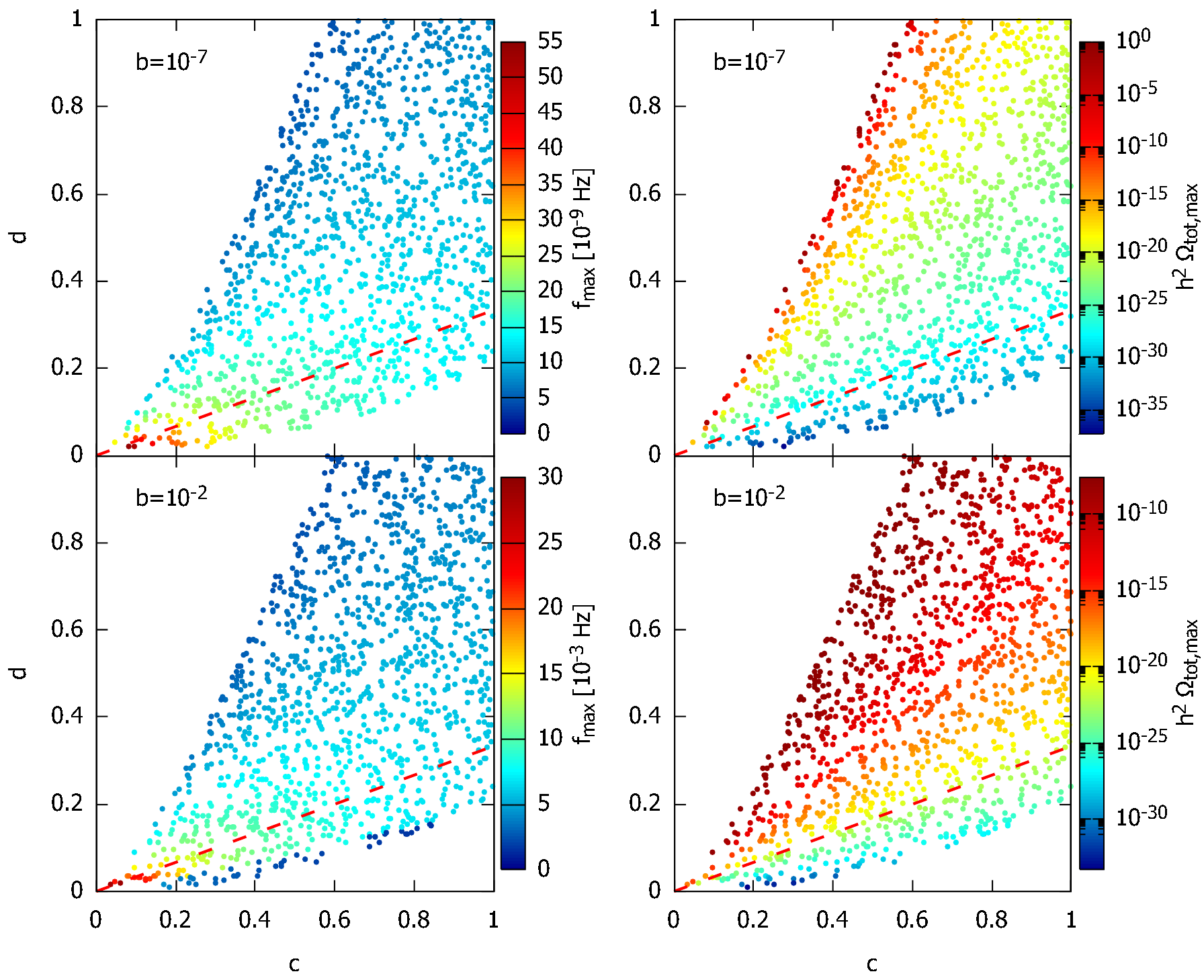}}
    \caption{GW signal strength $\Omega_{\text{tot,max}}$ (right) and peak frequency $f_{\text{max}}$ (left) in the $(c,d)$ plane for $b=10^{-7}$ (top) and $b=10^{-2}$ (bottom). The red dashed line ($d = c/3$) marks the boundary for a strong first-order phase transition (see eq.~(\ref{3-12})). Note the vastly different scales on the color bars, indicating that the low-$b$ regime can produce signals ruled out by cosmological observations.
    } \label{fig7}
\end{figure}

The results of our extensive parameter scan are visualized in Figure~\ref{fig7}, which provides a comprehensive overview of the generated GW phenomenology. The four-panel plot contrasts the two distinct regimes defined by the parameter $b$. For the smaller value, $b = 10^{-7}$ (top row), the peak frequencies $f_{\text{max}}$ are clustered on the order of nanohertz (left), positioning the potential signal in the frequency band targeted by pulsar timing arrays (PTA). Conversely, for $b = 10^{-2}$ (bottom row), the frequencies shift to the millihertz range, relevant for space-based interferometers like LISA.

The most striking feature, however, is the dramatic difference in the predicted signal strength $\Omega_{\text{tot,max}}$ (right column). The low-$b$ regime ($b=10^{-7}$) produces extraordinarily strong signals, with energy densities reaching $h^2\Omega_{\text{tot,max}} \sim 1$. In contrast, the high-$b$ scenario ($b=10^{-2}$) yields signals that are many orders of magnitude weaker, peaking around $h^2\Omega_{\text{tot,max}} \sim 10^{-10}$.

This result must be immediately contextualized within the framework of existing experimental and cosmological constraints. The amplitude of a stochastic GW background is severely bounded by both direct detection experiments and the success of standard cosmology. The most stringent constraints come from:
\begin{itemize}
    \item Big Bang Nucleosynthesis (Integrated bound): The total energy density in GWs present at the time of BBN cannot be so large as to alter the predicted abundances of light elements. This places a stringent integrated constraint \cite{Caprini:2018mtu}: 
\begin{equation*}
\int d(\ln f) \, h^2\Omega_{\text{GW}}(f) \lesssim 10^{-6} .
\end{equation*}
    \item Pulsar Timing Arrays (Nanohertz band): Recent results have evidence for a background with $h^2\Omega_{\text{GW}} \sim 10^{-9}$, setting a firm upper limit for any \emph{additional} cosmological source in this frequency range \cite{NANOGrav:2023gor}.
\end{itemize}
Consequently, the portion of our parameter space predicting the most potent signals, specifically those points for $b=10^{-7}$ with $h^2\Omega_{\text{tot,max}} \gg 10^{-6}$, is robustly ruled out by observation. A GW signal with $h^2\Omega \sim 1$ would have dominated the energy budget of the universe, a scenario incompatible with the observed cosmological history.

Therefore, while some benchmark points in the low-$b$ regime are mathematically possible, they are not physically viable. The high-$b$ regime, however, remains highly phenomenologically relevant. Predictions of $h^2\Omega_{\text{tot,max}} \sim 10^{-10}$ in the millihertz band are not only permissible but also represent a prime target for future space-based interferometers like LISA \cite{LISA:2017pwj, Caprini:2019egz}.

\begin{figure}[H]
\centerline{\hspace{0cm}\epsfig{figure=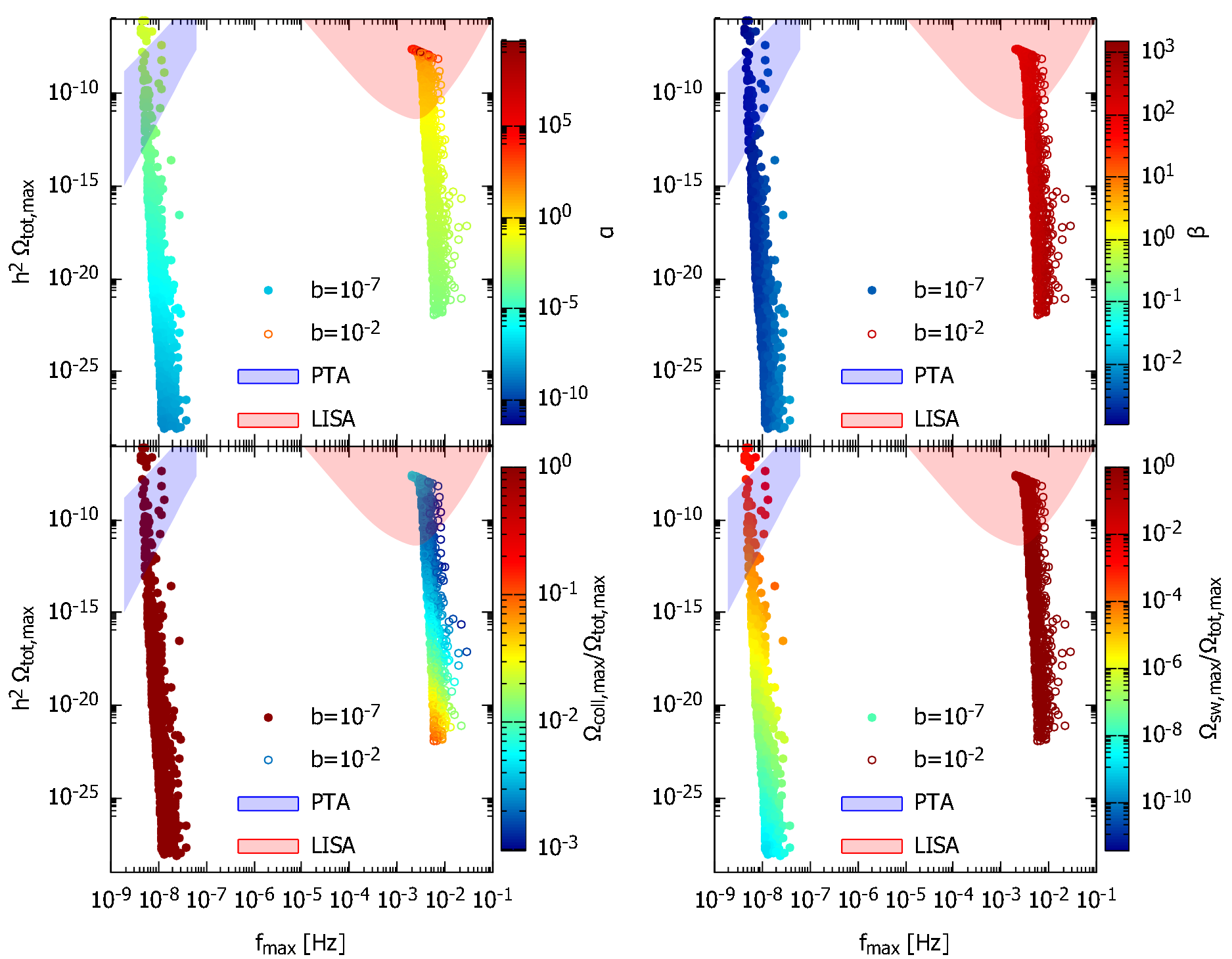,width=13cm}}
\caption{Correlations between GW observables and phase transition parameters for the viable parameter space. The total energy density is plotted against the peak frequency, with the points colored by: (Top Left) the transition strength $\alpha$, (Top Right) the inverse duration $\beta$, (Bottom Left) the fractional contribution from bubble collisions, and (Bottom Right) the fractional contribution from sound waves. Filled circular points denote $b=10^{-7}$ and hollow circular points $b=10^{-2}$. The shaded regions indicate the approximate sensitivity curves for Pulsar Timing Arrays (PTA, blue) and the LISA mission (red). The $b=10^{-2}$ signals, with their millihertz peak frequencies, are prime targets for the LISA mission. Conversely, the $b=10^{-7}$ signals fall within the nanohertz band, making them potential targets for Pulsar Timing Arrays (PTA).
} \label{fig8}
\end{figure}

The correlation between the predicted GW signals and phase transition parameters is explored in figure~\ref{fig8}, which plots the total energy density $h^2\Omega_{\text{tot,max}}$ against the peak frequency $f_{\text{max}}$ for all viable parameter sets (with $ h^2\Omega_{\text{tot,max}}<10^{-6} $ and $ d>c/3 $). The colored points are further differentiated by the values of the transition strength $\alpha$ (top left), the inverse duration $\beta$ (top right), and the relative contributions from bubble collisions ($\Omega_{\text{coll,max}}/\Omega_{\text{tot,max}}$, bottom left) and sound waves ($\Omega_{\text{sw,max}}/\Omega_{\text{tot,max}}$, bottom right).

The most immediate conclusion from this figure is the stark experimental distinction between the two scenarios. The high-$b$ ($10^{-2}$) signals (hollow circular points) occupy a region of higher frequency ($f_{\text{max}} \sim 10^{-3}$ Hz). Crucially, a significant population of these points falls within the projected sensitivity band of the LISA mission, identifying them as prime candidates for future detection.
In contrast, the viable low-$b$ ($10^{-7}$) signals (filled circular points) are characterized by much lower peak frequencies ($f_{\text{max}} \sim 10^{-9}$ Hz). While a subset of these points falls within the PTA frequency band, for most points, their exceptionally low amplitudes place them far below the current sensitivity threshold of PTA experiments and likely beyond the reach of any foreseeable detector.

The distribution of points reveals important physical trends. The transition strength $\alpha$ (top left) is the primary driver of the signal amplitude, with larger $\alpha$ values (red points) generating stronger signals. The inverse duration $\beta$ (top right) shows a strong anti-correlation with the peak frequency, as expected from the scaling $f_{\text{max}} \propto \beta$. 

A pivotal result is the clear inversion of the dominant GW production mechanism between the two regimes, as shown in the bottom panels of Figure~\ref{fig8}. The low-$b$ parameter space ($b=10^{-7}$) is characterized by a signal dominated by the bubble collision contribution ($\Omega_{\text{coll,max}}/\Omega_{\text{tot,max}} \sim 1$). However, as $b$ increases to $10^{-2}$, the sound wave mechanism becomes the principal source, accounting for nearly the entire signal, while the collision contribution becomes negligible. This shift, combined with the trends in $\alpha$ and $\beta$, indicates that the parameter $b$ not only determines the intensity and frequency of the signal but also its fundamental physical origin. This figure provides a concise summary of the detectability and physical characteristics of the GW signals produced across our viable parameter space.

\section{Conclusion} \label{sec5}
In this study, we have conducted a thorough investigation of high-temperature vacuum decay and the associated generation of GWs within a general scale-invariant framework. By deriving the effective potential that incorporates tree-level, one-loop, finite-temperature, and Daisy resummation contributions, we reduced its high-temperature behavior to a function of three key parameters: $b$, $c$, and $d$. Our analysis demonstrated that the decay of the metastable vacuum is predominantly governed by thermal sphaleron transitions rather than quantum tunneling via instantons.

A pivotal outcome of our work is the identification of two distinct phenomenological regimes dictated by the parameter $b$, which sets the energy scale of the phase transition. For a low value of $b$ ($10^{-7}$), the phase transition produces an exceptionally strong GW background in the nanohertz frequency range. However, the immense amplitude of this signal ($h^2\Omega_{\text{GW}} \sim 1$) is incompatible with the successful history of Big Bang Nucleosynthesis and is thus ruled out as a viable cosmological scenario. For a high value of $b$ ($10^{-2}$), the transition yields a much weaker signal ($h^2\Omega_{\text{GW}} \sim 10^{-10}$) peaking in the millihertz band. This regime remains entirely viable and presents a prime target for future space-based GW detectors like LISA.

Furthermore, our parameter scan revealed a clear hierarchy in the peak frequencies ($f_{\text{coll}} < f_{\text{sw}} < f_{\text{turb}}$) and a shift in the dominant production mechanism. The low-$b$ regime is dominated by bubble collisions, while the high-$b$ regime is overwhelmingly dominated by sound waves. The transition strength $\alpha$ was confirmed as the primary driver of the signal amplitude, while the inverse duration $\beta$ controls the peak frequency.

Our model-independent analysis provides a clear mapping from the fundamental parameters of scale-invariant theories to their GW observables. It establishes robust criteria for model-building, showing that to produce a detectable yet cosmologically consistent signal, a model must realize restricted parameter space. The results serve as a valuable guide for constructing specific, phenomenologically successful scale-invariant models whose imprints on the cosmos may be revealed by the next generation of GW astronomy.

\providecommand{\href}[2]{#2}\begingroup\raggedright\endgroup

\end{document}